\documentclass[usenatbib,usegraphicx]{mn2e}
\usepackage{geometry}

\title[Statistics of pulsar polarization]
{Statistics of the individual-pulse polarization based on
propagation effects in pulsar magnetosphere}
\author[S. A. Petrova]{S. A. Petrova \thanks{E-mail:
petrova@ira.kharkov.ua}\\ Institute of Radio Astronomy, 4,
Chervonopraporna Str., Kharkov 61002, Ukraine}

\begin{document}



\maketitle


\begin{abstract}
Pulsar radio emission is modelled as a sum of two completely
polarized non-orthogonal modes with the randomly varying Stokes
parameters and intensity ratio. The modes are the result of
polarization evolution of the original natural waves in the hot
magnetized weakly inhomogeneous plasma of pulsar magnetosphere. In
the course of wave mode coupling, the linearly polarized natural
waves acquire purely orthogonal elliptical polarizations. Further
on, as the waves pass through the cyclotron resonance, they become
non-orthogonal. The pulse-to-pulse fluctuations of the final
polarization characteristics and the intensity ratio of the modes
are attributed to the temporal fluctuations in the plasma flow.

The model suggested allows to reproduce the basic features of the
one-dimensional distributions of the individual-pulse polarization
characteristics. Besides that, propagation origin of pulsar
polarization implies a certain correlation between the mode
ellipticity and position angle. On a qualitative level, for
different sets of parameters, the expected correlations appear
compatible with the observed ones. Further theoretical studies are
necessary to establish the quantitative correspondence of the
model to the observational results and to develop a technique of
diagnostics of pulsar plasma on this basis.
\end{abstract}



\section{Introduction}
Pulsar magnetosphere contains the ultrarelativistic
electron-positron plasma which streams along the open magnetic
lines. The radio emission observed from pulsars is undoubtedly
associated with the processes in the plasma flow, and its
characteristics are believed to reflect the properties of the
ambient plasma. Whatever the radio emission mechanism, it should
give rise to the natural waves of the plasma. The
ultrarelativistic strongly magnetized plasma of pulsars allows two
types of non-damping natural waves, the ordinary and extraordinary
ones, which are linearly polarized in orthogonal directions, in
the plane of the ambient magnetic field and perpendicularly to
this plane, respectively. Pulsar beam is believed to be an
incoherent mixture of the two polarization modes. Since the
electric vector of each mode is related to the magnetic field
direction, the position angle (PA) of linear polarization should
vary smoothly across the pulse as the pulsar beam rotates with
respect to an observer \citep{RC69}. The expected S-shaped swing
of PA is indeed observed in a number of pulsars. In other cases,
abrupt nearly orthogonal jumps of PA can break the smooth swing,
testifying to the presence of both polarization modes. In general,
these modes can be markedly non-orthogonal
\citep*[e.g.][]{G1,G2,G3}.

An extensive analysis of observational data has led to the
conclusion that the two polarization modes are simultaneously
present in pulsar radiation \citep{Mc98,Mc00}. From the
theoretical point of view, this is the only way to account for
partial depolarization of pulsar radiation, since the natural
waves are completely polarized by definition. Thus, the model of
superposed polarization modes with randomly varying intensities
has weighty observational and theoretical grounds. At the same
time, it faces serious difficulties and seems too simplified to
explain diverse and complicated behaviour of the individual pulse
polarization. Generally, the observed modes have elliptical rather
than linear polarization and can be non-orthogonal
\citep{KarI,Mc03}. Besides that, the observed pulse-to-pulse
polarization fluctuations cannot be solely attributed to the
variations of mode intensities - the polarization characteristics
of the modes should fluctuate as well \citep*[e.g.][]{KarIII}.
These difficulties can be avoided if one take into account
propagation effects during wave passage through the flow of pulsar
plasma. Then the fluctuations in the observed polarization
characteristics can be attributed to the fluctuations in the
parameters of the plasma. It should be noted that the propagation
origin of pulsar polarization implies a certain correlation
between the final polarization characteristics, since they are
determined by the instanteneous state of the plasma in the regions
of significant polarization evolution inside the pulsar
magntosphere. The indications of such a correlation are indeed
discovered in observations \citep{ES04,E04}.

Deep inside the magnetosphere, the plasma number density is so
large that the natural waves propagate in the regime of
geometrical optics, with the polarization planes being adjusted to
the local magnetic field direction. Further along the trajectory,
as the plasma density decreases considerably, the wave mode
coupling starts: Each of the incident waves becomes a coherent
mixture of the two natural waves peculiar to the ambient plasma.
As a result, the waves acquire some circular polarization and a
shift in PA, so that it no longer reflects the magnetic field
geometry. The analytical and numerical tracings of the
polarization evolution because of wave mode coupling have proved
that  at the conditions relevant to the magnetosphere of a pulsar
this process can be efficient enough to account for the typically
observed circular polarization and PA shift \citep{LP99,PL00,P03}.
For given plasma parameters, both types of waves acquire the same
shift in PA and the same fractional circular polarization of
opposite senses. Thus, the outgoing waves have the elliptical
polarization purely orthogonal at the Poincar\'e sphere. This
representation is compatible with the bulk of observational data.
However, sometimes a clear evidence is met that the polarization
modes are non-orthogonal \cite[e.g.][]{Mc03}.

The non-orthogonality of polarization modes can be attributed to
the cyclotron absorption in pulsar magnetosphere, if this process
follows the wave mode coupling \citep{P05,P06}. A similar idea has
recently been discussed in \citep{M06}. Typically the region of
cyclotron resonance lies beyond the region of efficient wave mode
coupling, but they can be quite close to each other, in which case
the plasma number density is still large enough for the resonance
to affect wave polarization considerably. For the two types of
natural waves, the rate of cyclotron absorption is slightly
different. Since the waves entering the resonance region present
already a coherent mixture of the two natural waves and these
constituents are absorbed not identically, the polarization state
of the waves changes. The original ordinary and extraordinary
waves suffer different polarization evolution and become
non-orthogonal.

The aim of the present paper is to study statistics of the
propagation-induced polarization of pulsars resulting from the
fluctuations in the plasma flow. Special attention is paid to
studying the correlation between the mode ellipticity and PA
because of propagation effects. Section~2 contains the general
equations of polarization evolution in pulsar plasma and an
example of numerical tracings of the ellipticity and PA. The
statistics of the final polarization parameters in the plasma with
randomly varying parameters are examined in Sect.~3. The
numerically simulated histograms of PA and ellipticity, as well as
the two-dimensional scatter plots of the Stokes parameters are
given. The results of the paper are discussed and summarized in
Sect.~4.

\section{General theory of polarization evolution in pulsar plasma}
Let the radio waves propagate through the ultrarelativistic
electron-positron plasma, which streams along the open field lines
of the dipolar magnetic field of a pulsar. Because of continuity
of the plasma flow within the open field line tube, the plasma
number density decreases with distance from the neutron star just
as the magnetic field strength, $N\propto B\propto z^{-3}$. Thus,
the radio waves propagate in the weakly inhomogeneous medium, with
the scale length for change in the parameters much larger than the
wavelength. The plasma is assumed to be hot, so that the region of
cyclotron resonance is also much larger than the wavelength.

The waves are believed to originate deep inside the magnetosphere,
where the plasma number density is large enough to provide the
conditions of geometrical optics and the magnetic field is strong
enough for the electron gyrofrequency to be much larger than the
radio frequency in the plasma rest frame. As both $N$ and $B$
decrease rapidly along the trajectory, both these conditions are
ultimately broken. First of all, the scale length for beats
between the natural waves, $L_b\sim\frac{c}{\omega\Delta n}$,
where $\Delta n$ is the difference in their refractive indices,
becomes comparable to the scale length for change in the medium
parameters, $L_p\sim z$, so that geometrical optics is violated
and wave mode coupling holds. In case of the cold
ultrarelativistic strongly magnetized plasma,
\begin{equation}
\Delta n=\frac{2\omega_p^2}{\gamma^3\omega^2\theta^2}\,,
\end{equation}
where $\omega_p\equiv\sqrt{\frac{4\pi Ne^2}{m}}$ is the plasma
frequency, $\gamma$ the plasma Lorentz-factor, $\theta$ the
wavevector tilt to the ambient magnetic field and it is taken into
account that the wave propagation is generally quasi-transverse,
$\theta\gg 1/\gamma$. Then the characteristic radius of wave mode
coupling is determined by the relation
\begin{equation}
\frac{2\omega_p^2(r_p)r_p}{\gamma^3c\omega\theta^2(r_p)}=1\,.
\end{equation}
In case of hot plasma, equations (1) and (2) remain the same, but
$\gamma$ means some characteristic Lorentz-factor of the plasma
particle distribution. For the conditions relevant to pulsar
magnetosphere, the radius of wave mode coupling, $r_p$, can be
estimated as \citep[equation (16) in][]{P06}:
\begin{equation}\frac{r_p}{r_L}=0.18P^{-3/2}\gamma_{1.5}^{-3/2}\kappa_2^{1/2}B_{\star
12}^{1/2}\nu_9^{-1/2}\theta_{-1}^{-1}\,.
\end{equation}
Here $r_L$ is the light cylinder radius, $P$ the pulsar period,
$\kappa$ the plasma multiplicity,
$\kappa_2\equiv\frac{\kappa}{10^2}$, $B_\star$ the magnetic field
strength at the surface of the neutron star,
$B_{\star_{12}}\equiv\frac{B_\star}{10^{12}\,{\rm G}}$, $\nu$ the
radio frequency, $\nu_9\equiv\frac{\nu}{10^9\,{\rm Hz}}$,
$\gamma_{1.5}\equiv\frac{\gamma}{10^{1.5}}$,
$\theta_{-1}\equiv\frac{\theta}{10^{-1}}$. One can see that $r_p$
is determined by the basic pulsar parameters, $P$ and $B_\star$,
as well as by the parameters of the plasma flow, $\kappa$ and
$\gamma$. Besides that, $r_p$ strongly depends on the angle
$\theta$, which introduces a significant uncertainty in the
estimate (3). The value of $\theta$ may differ within an order of
magnitude ($\theta\sim 0.1-1$) for different pulsars and for the
rays observed at different longitudes in a given pulsar. On the
whole, one can conclude that in the majority of pulsars for all
the rays forming the pulse profile $r_p$ lies inside the light
cylinder.

It should be noted that the wave mode coupling holds only if the
wavevector goes out of the initial plane of magnetic lines. In
pulsar case, this condition is certainly satisfied, e.g. because
of the magnetosphere rotation. Let us choose the Cartesian
coordinate system with the z-axis along the wavevector and the
x-axis in the initial plane of magnetic lines (see Fig.~1). The
ray is emitted along a field line of the dipolar magnetic field at
an angle $\psi$ to the magnetic axis. In the non-rotating
magnetosphere, the ray would propagate in the plane of magnetic
lines and at distances much larger than the emission altitude
would make the angle $\sim\psi$ with the ambient magnetic field
because of divergence of the magnetic lines, so that
$b_x\sim\psi$, $b_y=0$. Because of the magnetosphere rotation,
$b_y$ increases with distance $\propto z/r_L$ (the exact formulas
for $b_x$ and $b_y$, which allow for the rotational aberration,
are given by the equation (3) in \citet{P03}). The quantity
$\mu\equiv(b_y/b_x)_{r_p}\sim r_p/(r_L\psi)$ is the key parameter
which determines the efficiency of wave mode coupling. It should
be noted that although as a rule $r_p/r_L\ll 1$, $\mu$ can be of
order unity and even larger. The process of wave mode coupling is
most efficient at $\mu =1$, in which case the resultant degree of
circular polarization of the waves is as large as about 40\% (cf.
Fig.~5 in \citet{P01}). Generally speaking, $\mu$ depends on the
geometry of ray propagation and can differ substantially for the
rays observed at different pulse longitudes and frequencies as
well as for different pulsars. Moreover, $r_p$ and $\mu$ may vary
from pulse to pulse because of fluctuations in the plasma
parameters (cf. equation (3)).

The process of wave mode coupling holds as long as $L_b\sim L_p$.
As the plasma number density decreases further, this process
ceases and the waves propagate as in vacuum, preserving their
polarization state. For this reason, the wave mode coupling is
usually called polarization-limiting effect. However, actually it
is the final stage of polarization evolution only if the region of
cyclotron resonance lies infinitely far. Otherwise the cyclotron
absorption also affects the wave polarization. For the particles
with the characteristic Lorentz-factor, the condition of cyclotron
resonance, $\omega\gamma\theta^2/2=\omega_H\equiv\frac{eB}{mc}$,
is met at
\begin{equation}
\frac{r_c}{r_L}=0.55P^{-1}B_{\star
12}^{1/3}\nu_9^{-1/3}\gamma_{1.5}^{-1/3}\theta_{-1}^{-2/3}
\end{equation}
\citep[e.g.][]{P06}. Similarly to the radius of wave mode
coupling, the resonance radius is strongly affected by the
uncertainty in the angle $\theta$, but generally $r_c$ lies within
the light cylinder. Comparing the equations (3) and (4), one can
conclude that for a wide range of pulsar parameters $r_c\sim r_p$
and the process of cyclotron absorption can play a marked role in
determining the final polarization of radio waves. The quantity
$\eta\equiv r_p/r_c$ is another key parameter of the polarization
evolution in pulsar magnetosphere.

Generally speaking, the plasma of pulsar magnetosphere is
gyrotropic, i.e. the distribution functions of electrons and
positrons are somewhat different. Correspondingly, the natural
waves are linearly polarized only in the vicinity of the emission
region, where the approximation of infinitely strong magnetic
field is still valid. In the outer magnetosphere, in the regions
of significant polarization evolution (at $r\sim r_p,r_c$) the
natural waves of the medium are generally elliptical.
Unfortunately, the question on the net current density and its
distribution in the magnetosphere is still open. Given that the
electrons and positrons differ only in the number densities and
this difference is of the order of the Goldreich-Julian number
density, i.e. $\Delta N/N\sim 1/\kappa$, the relative contribution
of the plasma gyrotropy to the polarization evolution in the
region of wave mode coupling can be estimated as
$0.1\kappa_2^{-1}\theta_{-1}^2\gamma_{1.5}^2\omega^\prime
/\omega_H$ \citep{P06}; here $\omega^\prime$ is the wave frequency
in the plasma frame, $\omega^\prime\equiv\omega\gamma\theta^2/2$.
One can see that if the wave mode coupling holds close enough to
the region of cyclotron resonance,i.e. if $\omega^\prime
/\omega_H\sim 1$, the contribution of the intrinsic ellipticity of
the natural waves can be significant, at least for the rays of a
specific geometry. However, as the true form of the net charge
density is unknown, we leave out the plasma gyrotropy, keeping in
mind that this effect can also be significant and may somewhat
alter our quantitative results. In the present paper, we
concentrate on studying the evolution of linearly polarized
natural waves as a result of propagation effects in pulsar plasma.
In our consideration, the wave ellipticity arises and changes
purely on account of wave mode coupling and cyclotron absorption.
An opposite approach has recently been developed by \citet{LM04}
and \citet{ML04a}, who investigated the characteristics of the
elliptically polarized natural waves at $r\sim r_p$, ignoring the
wave mode coupling in this region. The effect of cyclotron
absorption on the elliptical natural waves has been considered in
\citet{ML04b}.

An exact treatment of polarization transfer in pulsar plasma with
account for the wave mode coupling and cyclotron absorption can be
found in \citet{P06}. The evolution of the Stokes parameters of
the natural waves is described by the following set of equations:
\[\frac{{\rm d}I}{{\rm d}w}=-\eta^2swF_2q\left (1-\frac{\mu^2}{\eta^2w^2}\right )
-2\mu\eta sF_2u\,,\]
\[\frac{{\rm d}q}{{\rm d}w}=-\eta^2swF_2I\left (1-\frac{\mu^2}{\eta^2w^2}\right )
-2\mu\eta sF_1v\,,\]
\[\frac{{\rm d}u}{{\rm d}w}=\eta^2swF_1v\left (1-\frac{\mu^2}{\eta^2w^2}\right )
-2\mu\eta sF_2I\,,\]
\begin{equation}
\frac{{\rm d}v}{{\rm d}w}=-\eta^2swF_1u\left
(1-\frac{\mu^2}{\eta^2w^2}\right ) -2\mu\eta sF_1q\,.
\end{equation}
Here $w\equiv r_c/z$,
$s\equiv\frac{1+\mu^2}{(1+\mu^2/(\eta^2w^2))^2}$,
\[F_1\equiv {\rm v.p.}\int \frac{f(\gamma){\rm
d}\gamma}{(\gamma /\gamma_0)^3[1-(\gamma /\gamma_0)^2w^{-6}]}\,,\]
\[F_2\equiv\frac{\pi}{2}w^{-6}\gamma_0f\left
[w^{3}\gamma_0\right ]\,,\] v.p. means that the integral is taken
in the principal value sense, $f(\gamma)$ is the particle
distribution function with the normalization $\int f(\gamma){\rm d
}\gamma =1$, $\gamma_0$ the characteristic Lorentz-factor of the
plasma particles, $r_c$ the characteristic radius of cyclotron
resonance defined as
$\omega_H(r_c)\equiv\omega\gamma_0\theta^2/2$. In equation (5) we
have omitted the factors common to all the Stokes parameters,
since they are irrelevant to the problem of polarization evolution
considered. \footnote{Note, however, that the cyclotron absorption
can markedly affect the total intensity of pulsar radiation. In
the short-period pulsars, $P\sim 0.1$ s, the optical depth to this
process can exceed unity \citep{LP98}. As is shown in \citet{P02},
the effect of resonant absorption can account for the observed
statistical features in the energetic characteristics of the
short-period pulsars.} The set of equations (5) is a
generalization of equations (17) in \citet{P06} for the case of
arbitrary $\mu$. At the same time, we still do not include the
terms responsible for rotational aberration explicitly and assume
that they enter $b_x$ and $b_y$ as factors of order unity (for
more detail see \citealt{P06}). The terms containing $F_1$
describe polarization evolution because of wave mode coupling,
while those containing $F_2$ correspond to cyclotron absorption.
The initial conditions for the natural waves read:
$(I,q,u,v)_0=(1,\pm 1,0,0)$, where the upper and lower signs refer
to the ordinary and extraordinary waves, respectively.

To proceed further, we choose a simplified distribution function
of the plasma particles in the form of a triangle:
\begin{equation}
f(\gamma)=\left\{\begin{array}{lr}\frac{\displaystyle\gamma
-\gamma_0(1-a) }{\displaystyle
a^2\gamma_0^2}\,,&\quad\gamma_0(1-a)\leq\gamma\leq\gamma_0\,,\\
\frac{\displaystyle\gamma_0(1+a)-\gamma}{\displaystyle
a^2\gamma_0^2}\,,&\quad\gamma_0\leq \gamma\leq\gamma_0(1+a)\,.
\end{array}\right.
\end{equation}
Generally speaking, the numerical results on polarization
evolution are not very sensitive to the detailed shape of
$f(\gamma)$ and are mainly affected by the average Lorentz-factor,
$\gamma_0$, and the width of the distribution (the parameter $a$
in equation (6)). The set of equations (5) along with equation (6)
describe polarization evolution of the original natural waves in
the hot magentized weakly inhomogeneous plasma of pulsars.

Since the natural waves are completely polarized by definition,
their polarization state is described by two independent
quantities, e.g. PA, $\psi\equiv 0.5\arctan (u/q)$, and the
ellipticity, $\chi\equiv 0.5\arctan\frac{v}{\sqrt{q^2+u^2}}$.
Figure~2 shows an example of numerical tracings of $\psi$ and
$\chi$ of the original ordinary and extraordinary waves as they
propagate in pulsar magnetosphere. One can see that polarization
evolution can indeed be significant and differs markedly for the
two modes.

\section{Statistics of the final polarization parameters}
\subsection{Histograms of PA and ellipticity}
As is discussed above, polarization evolution is determined by the
parameters $\mu$ and $\eta$ and is affected by the parameters of
the particle distribution function. For the rays observed at a
fixed pulse longitude, the locations of the regions of wave mode
coupling and cyclotron resonance may vary from pulse to pulse
because of fluctuations in the number density and the
characteristic Lorentz-factor of the plasma particles. Besides
that, the fluctuations in the plasma distribution may affect
refraction of the waves, so that the rays of a certain orientation
may follow somewhat different trajectories in the open field line
tube. All this is believed to influence the final polarization
states of the original natural waves. Note that although the
modern theories of the pulsar pair creation cascade say nothing
about the fluctuations in the resultant distribution of the
secondary plasma, an idea of such fluctuations is supported by the
random nature of the cascade process and is strongly suggested by
the variability of the observed individual pulse profiles.

In the upper panel of Fig.~2, the left plot shows the histogram of
the final PA of the original natural waves in case of randomly
varying parameters $\mu$ and $\eta$. One can see that the peaks
are separated by not exactly $90^\circ$, the scatter of PA values
around the peaks is about $10^\circ$ and the ordinary mode (that
with positive PA) shows somewhat less scatter than the
extraordinary one. The histogram of mode ellipticities for the
same $\mu$ and $\eta$ is shown in the right plot of the upper
panel in Fig.~2. Note that the peaks are not symmetrical with
respect to zero and the extraordinary mode exhibits a more
pronounced scatter.

The middle and bottom panels of Fig.~3 show the histograms of PA
and ellipticity for the sum of the two polarization modes given
random variations in $\mu$, $\eta$ and the mode intensity ratio.
In the present consideration, we assume that initially only the
ordinary mode is emitted, whereas the extraordinary one arises as
a result of partial conversion  of the ordinary mode deep inside
the magnetosphere \citep[see][]{P01}. This is supported by the
recently discovered anticorrelation of the mode intensities
\citep{ES04}. Since the mode conversion is a propagation effect,
its efficiency is also determined by the instantaneous
distribution of the plasma and is believed to fluctuate. Thus, the
coefficient of conversion, $\tau$, is considered as a random
quantity. Generally speaking, it may be correlated with $\mu$ and
$\eta$, but as the process of mode conversion takes place far from
$r_p$ and $r_c$, the correlation is thought to be weak and is
neglected throughout the paper. In the middle panel of Fig.~2,
$\tau$ is taken to be distributed over the whole interval from 0
to 1, whereas in the bottom panel $\tau\in [0.6,1]$.

Although in the former case the probability to dominate is the
same for the two types of original natural waves, the humps at the
PA histograms are substantially distinct: the peak corresponding
to the ordinary waves is much more pronounced. This is a
consequence of cyclotron absorption, which suppresses the
extraordinary constituent of the wave polarization stronger. The
extraordinary-mode hump looks more smeared and is connected to the
ordinary-mode one with a bridge. In the bottom panel, the humps
change in dominance, but the ordinary mode is still present,
despite complete dominance of the extraordinary waves just after
conversion. This is again because of differential action of
cyclotron absorption on the two types of natural waves. The bridge
between the humps looks more pronounced, and the histogram on the
whole is similar to the observed ones \citep[e.g.][]{Mc03}.

In the middle histogram of the ellipticity, the hump corresponding
to the extraordinary mode is barely resolved, whereas the ordinary
mode peaks at $\chi\approx 0$. On the whole, this distribution
looks like a unimodal one with a long tail. As can be seen in the
bottom histogram of $\chi$, only positive values are met and the
two humps are barely resolved. This distribution can also be
regarded as a unimodal one, in contrast to the corresponding
histogram of PA. It is not our aim here to fit the concrete
distributions observed, but Fig.~3 demonstrates the principal
possibility of such fits within the framework of the propagation
model of pulsar polarization.

\subsection{Two-dimensional scatter-plots}
Propagation origin of pulsar polarization implies a certain
correlation between the mode ellipticity and PA: Both these
quantities are determined by the instantaneous state of the plasma
and vary from pulse to pulse because of fluctuations in the plasma
flow. Now we are going to model the observational signatures of
this correlation for different sets of $\mu$ and $\eta$ and
compare our simulations with the observational results.

At a fixed pulse longitude, the simulated individual-pulse
polarization presents a sum of the two polarization modes with
random intensity ratio. The resultant Stokes parameters are
plotted in the Lambert's azimuthal equal area projection, as is
done by observers \citep{ES04,E04}. As each mode can dominate from
time to time, at the Poincar\'e sphere the resultant Stokes
vectors tend to group around the roughly orthogonal locations
corresponding to the observed modes. Let us choose the spherical
coordinate system with the polar axis along the fiducial vector of
the normalized Stokes parameters, ${\bmath s}_p=\{q_p,u_p,v_p\}$,
$q_p^2+u_p^2+v_p^2=1$, which is the average over the Stokes
vectors belonging to one of the observed modes. The quantities
$\rho$ and $\lambda$ are the polar angle and azimuth of the
normalized individual Stokes vectors (q,u,v) in this spherical
system:
\[\rho=\arccos (qq_p+uu_p+vv_p)\,,\]
\begin{equation}
\lambda=\arctan\left
[\frac{uq_p-qu_p}{(qq_p+uu_p)v_p-v(q_p^2+u_p^2)}\right ]\,.
\end{equation}
If one consider $\rho$ and $\lambda$ as the polar coordinates (the
radius and azimuth, respectively), we come to the Lambert's
azimuthal equal area projection of the Poincar\'e sphere. This
projection is interrupted at the equator, and for the Stokes
vectors lying in the southern hemisphere (i.e. for the second
mode) the projection of the sphere with the polar axis along
${\bmath s}_m=-{\bmath s}_p$ is considered \citep[for more detail
see][]{ES04}.

Figure~4 shows the two-dimensional plots for the two original
natural modes after their evolution in the plasma in case of
negligible cyclotron absorption, $\eta\sim 0.01$, and moderate
mode coupling, $\mu\sim 0.1$. Here ${\bmath s}_p$ corresponds to
the original ordinary mode. One can see that the process of mode
coupling in the fluctuating plasma results in the unambiguous
relation of the polarization characteristics, which has already
been proposed as a basis for diagnostics of pulsar plasma
\citep{P03}. At these plots, the points prefer certain azimuths,
which is in qualitative agreement with the observational result of
\citet{E04} (see Fig.~5 there), though the scatter of the
observational points is enormously large. Note that this scatter
cannot be reproduced by considering the sum of the two modes with
random intensities. Introducing a slight non-orthogonality of the
modes because of cyclotron absorption ($\eta\sim 0.1$) makes the
plot more realistic (see Fig.~5 for the case of completely
dominating original ordinary mode, $\tau\in [0,0.3]$), though the
scatter of the  points is still insufficient.

Figure~6 shows the two-dimensional scatter plots in case of strong
non-orthogonality of the modes ($\eta\sim 1$). In the upper panel,
the fiducial Stokes vector, ${\bmath s}_p$, corresponds to the
state with the dominant ordinary mode, whereas the coefficient of
conversion is uniformly distributed in the interval $[0.5,1]$.
Note that although in this case the conversion of the ordinary
waves into the extraordinary ones is efficient, the final
polarization state of the sum of the modes can still be close to
that of the ordinary ones because of less significant absorption
of the ordinary waves and because of a marked non-orthogonality of
the modes. The bottom panel of Fig.~6 corresponds to the case of a
uniform distribution of $\tau$ over the whole interval from 0 to
1, with ${\bmath s}_p$ directed along the extraordinary mode
vector. The plots in both panels exhibit strong non-orthogonality
of polarization states, with the secondary modes forming the
characteristic arc-like features, which are in a qualitative
agreement with those discovered in observations \citep[cf. Fig.~2
in ][]{ES04}. Note, however, that the scatter of the simulated
points and the degree of non-orthogonality are substantially less
than those in the observational plots. This may be an indication
of the role of some additional effects in the formation of the
single-pulse polarization of pulsars.

\section{Discussion and conclusions}
The hot magnetized weakly inhomogeneous plasma of pulsars may
substantially affect the radio wave propagation. The polarization
states of the original natural waves may change markedly because
of the wave mode coupling and cyclotron absorption. The former
process turns the original linearly polarized waves into the
elliptical ones, which are still purely orthogonal at the
Poincar\'e sphere. The mode coupling efficiency is determined by
the location of the coupling region, $r_p$, in the tube of open
magnetic lines. The role of cyclotron absorption in the evolution
of wave polarization depends on how close the regions of mode
coupling and cyclotron resonance are, being most prominent in case
of their approximate coincidence, $\eta=r_p/r_c\approx 1$.
Typically $r_c$ lies beyond $r_p$, in which case cyclotron
absorption leads to the non-orthogonality of the waves. Note that
all the way in the magnetosphere the original natural waves remain
completely polarized.

Temporal fluctuations in the plasma flow are believed to underlie
the fluctuations of the individual pulse polarization. The plasma
distribution in the open field line tube can be strongly variable
because of the random character of the pair creation process.
Hence, for the rays observed at a fixed pulse longitude, both
$r_p$ and $r_c$ can change from pulse to pulse due to variations
of the number density and characteristic Lorentz-factor of the
plasma. Besides that, the variations of the plasma distribution
may affect refraction of waves, so that the rays observed at a
given pulse longitude may follow somewhat different trajectories
in the magnetosphere and have different tilts to the ambient
magnetic field while passing through the regions of mode coupling
and cyclotron resonance.

The intensity ratio of the modes fluctuates as well, and it is
also thought to result from the plasma fluctuations in the
magnetosphere. The point is that the extraordinary natural waves
have vacuum dispersion and can hardly be generated directly by any
conceivable emission process in the plasma. Therefore they are
likely to originate as a result of partial conversion of the
ordinary waves, which can take place in the regions of
quasi-longitudinal propagation deep inside the magnetosphere
\citep{P01}. The idea of mode conversion is supported by the
recently discovered anticorrelation of the mode intensities
\citep{ES04}. In the framework of this view, the mode intensity
ratio is determined by the coefficient of conversion and changes
from pulse to pulse because of fluctuations in the plasma flow.

Thus, the propagation model of pulsar polarization incorporates
two superposed completely polarized modes with randomly varying
polarization states and intensities. It is important to note that
because of cyclotron absorption the superposed modes can become
non-orthogonal, whereas the original natural waves are purely
orthogonal by definition.

In the present paper, the statistics of the individual pulse
polarization have been simulated under the assumption that the
parameters $\mu$ and $\eta$ as well as the coefficient of
conversion, $\tau$, are the random quantities with unifirm
distributions over some intervals. The resultant histograms of PA
exhibit two humps markedly smeared and connected with a bridge.
The peaks can be separated by not exactly $90^\circ$. The
histograms of the resultant ellipticity look like the unimodal
ones because of a very small separation between the peaks of the
two observational modes. Furthermore, it may happen that in the
whole sample the ellipticity is purely of one sign, whereas the
corresponding histogram of PA is bimodal. Although direct fits to
the observational data are beyond the scope of the present paper,
our results confirm the ability of the propagation model to
account for the main features of the observed histograms of PA and
ellipticity.

Propagation origin of pulsar polarization implies a certain
correlation between the mode ellipticity and PA. Given that the
contribution of cyclotron absorption is negligible and the final
polarization is determined solely by the mode coupling, there is a
one-to-one correspondence between the mode ellipticity and PA.
However, it is not proved by the observational data available.
Although an evidence for the expected relation is indeed present
in the polarization of PSR B0818-13 \citep{E04}, the scatter of
the observational points appears dramatic, and it cannot be
reproduced by taking into account the fluctuations of the mode
intensity ratio. In case of moderately weak cyclotron absorption,
a slight non-orthogonality of the fluctuating modes causes an
additional scatter of the polarization parameters and allows to
reproduce the observational plots on a qualitative level, though
the scatter is still insufficient quantitatively. In case of
strong non-orthogonality of the modes, our model qualitatively
reproduces the characteristic arc-like features present in the
observational plots for PSR B0329+54 \citep{ES04}, though the
magnitude of non-orthogonality and the total scatter of the
simulated points are again less than in observations.

Thus, on a qualitative level, the expected correlations of
polarization parameters are compatible with the observational
data. This is a strong argument in favour of the propagation model
of pulsar polarization. Apparently, to achieve better quantitative
agreement with the observational results it is necessary to
improve and further develop the model suggested. In the present
consideration, we have ignored the net charge density in the
magnetosphere and treated the process of cyclotron absorption in
the limit of small pitch-angles of the particles. In reality, both
these assumptions can be violated, so that our results on the
individual-pulse polarization and its statistics can be somewhat
modified. Besides that, pulsar emission can be considered as a sum
of contributions from multiple subsources \citep[e.g.][]{G3,M06},
in which case the possibilities of modeling the resultant
polarization are much wider.

In conclusion, it should be pointed out that the model suggested
allows a unique possibility of diagnostics of pulsar plasma by
means of the individual-pulse polarization.

\section*{Acknowledgements}
This research is in part supported by INTAS Grant No.~03-5727 and
the Grant of the President of Ukraine (the project No.~GP/F8/0050
of the State Fund for Fundamental Research of Ukraine).

\input epsf

\clearpage
\begin{figure*}
\includegraphics[width=130mm]{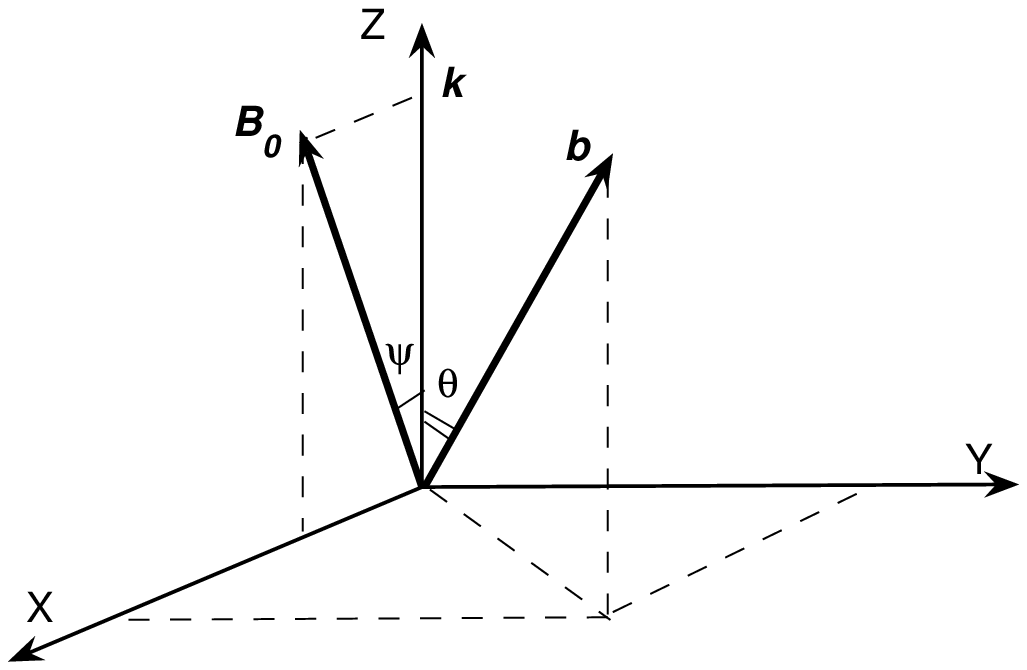}
\caption[]{Geometry of the vectors $\bmath k$ and $\bmath b$ in
the coordinate system described in the text. $\bmath {B_0}$ is the
initial orientation of the magnetic axis.}
\end{figure*}

\begin{figure*}
\includegraphics[width=100mm]{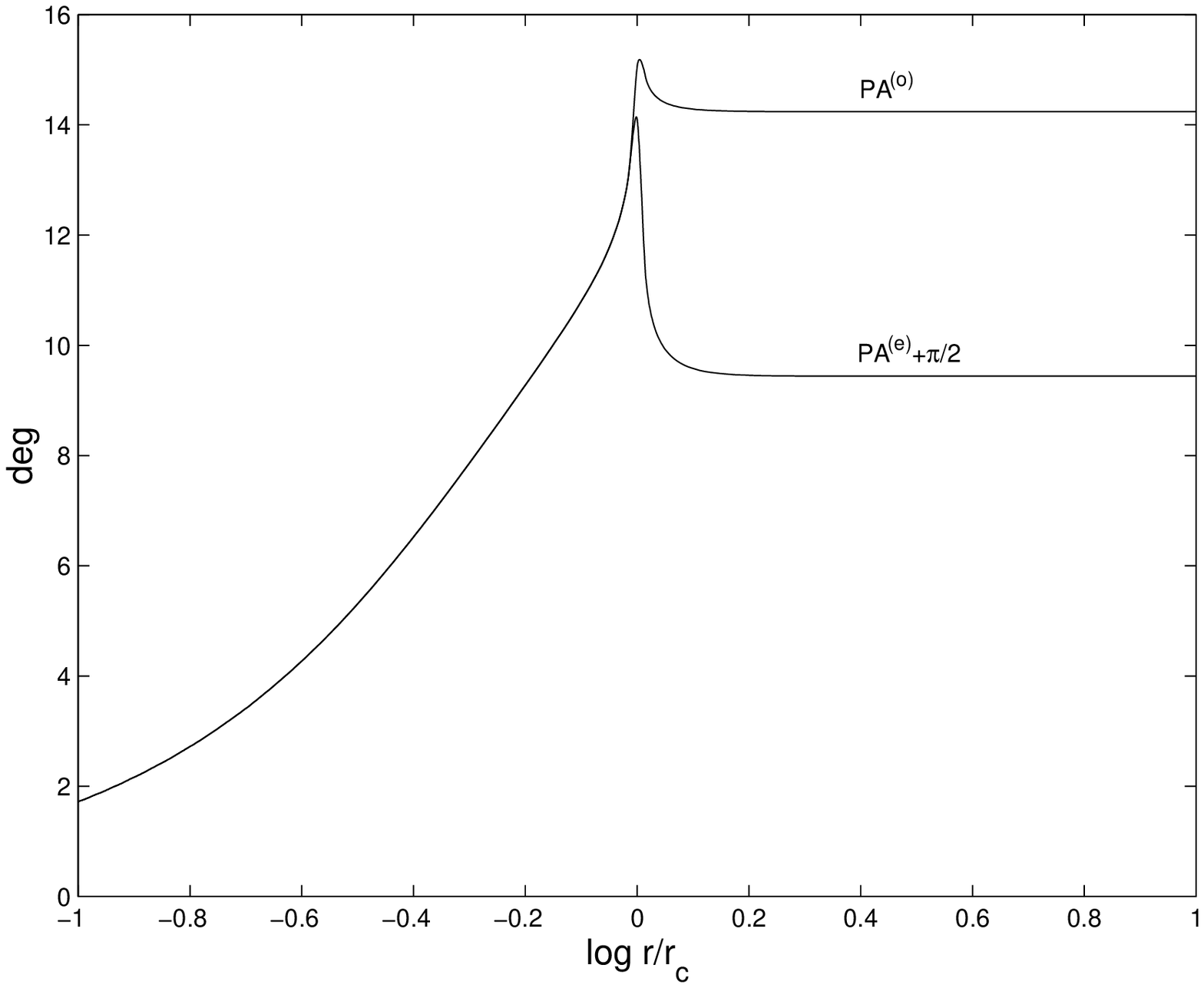}
\includegraphics[width=100mm]{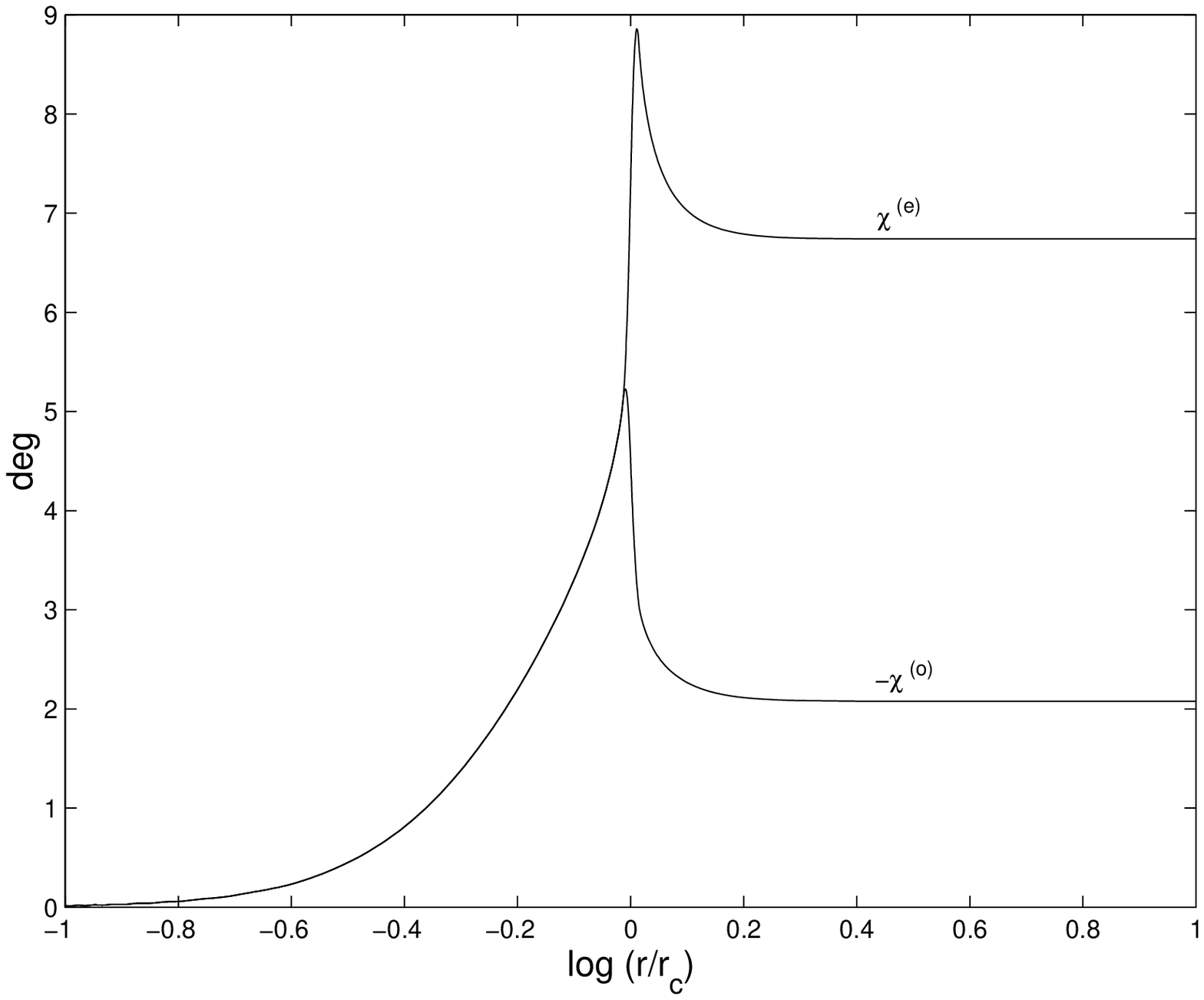}
\caption[]{Position angle ({\bf a}) and ellipticity ({\bf b}) of
the original ordinary and extraordinary modes as functions of
distance in pulsar magnetosphere; $\mu=0.3$, $\eta=1$, $a=0.1$.}
\end{figure*}

\begin{figure*}
\includegraphics[width=60mm]{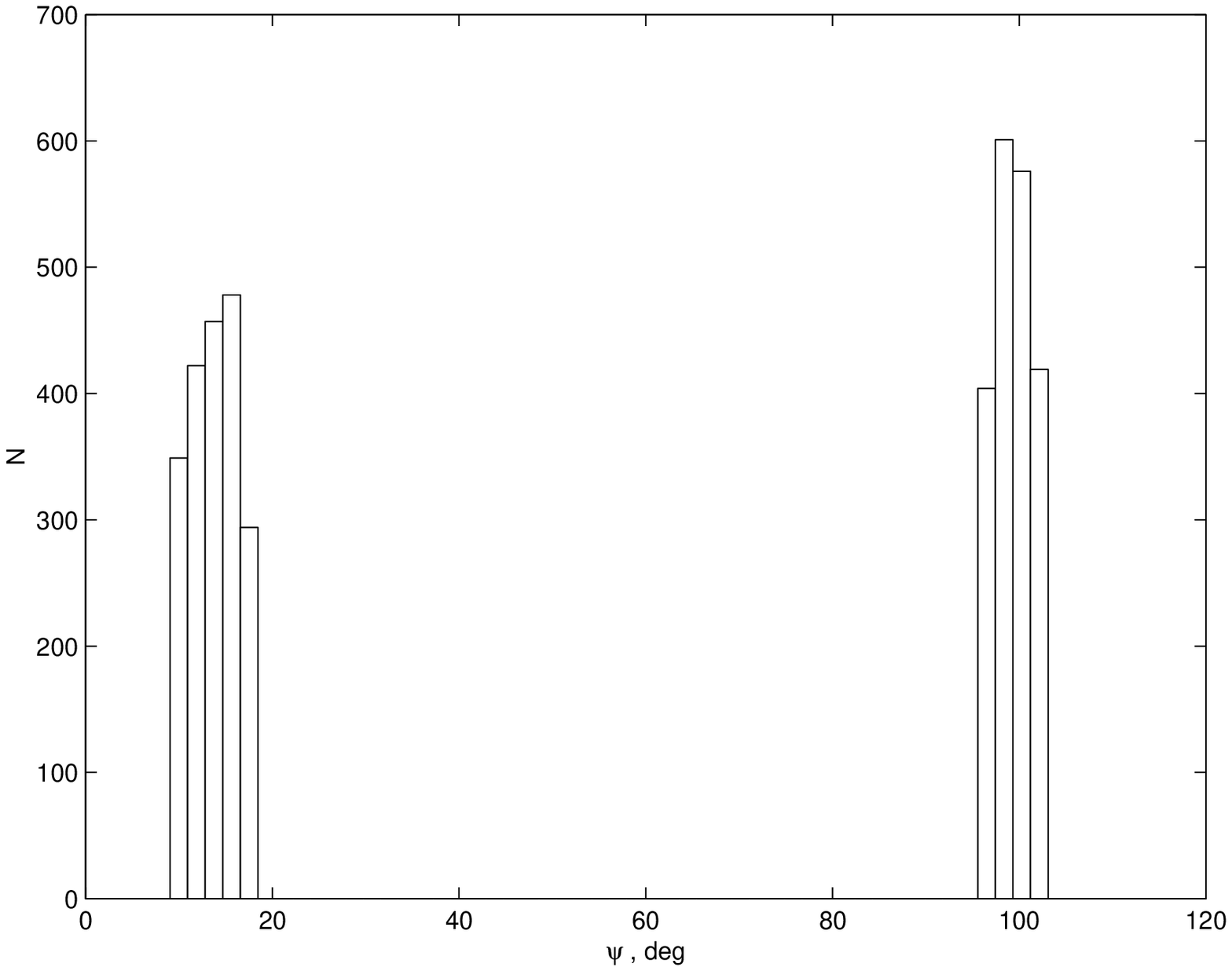}\includegraphics[width=60mm]{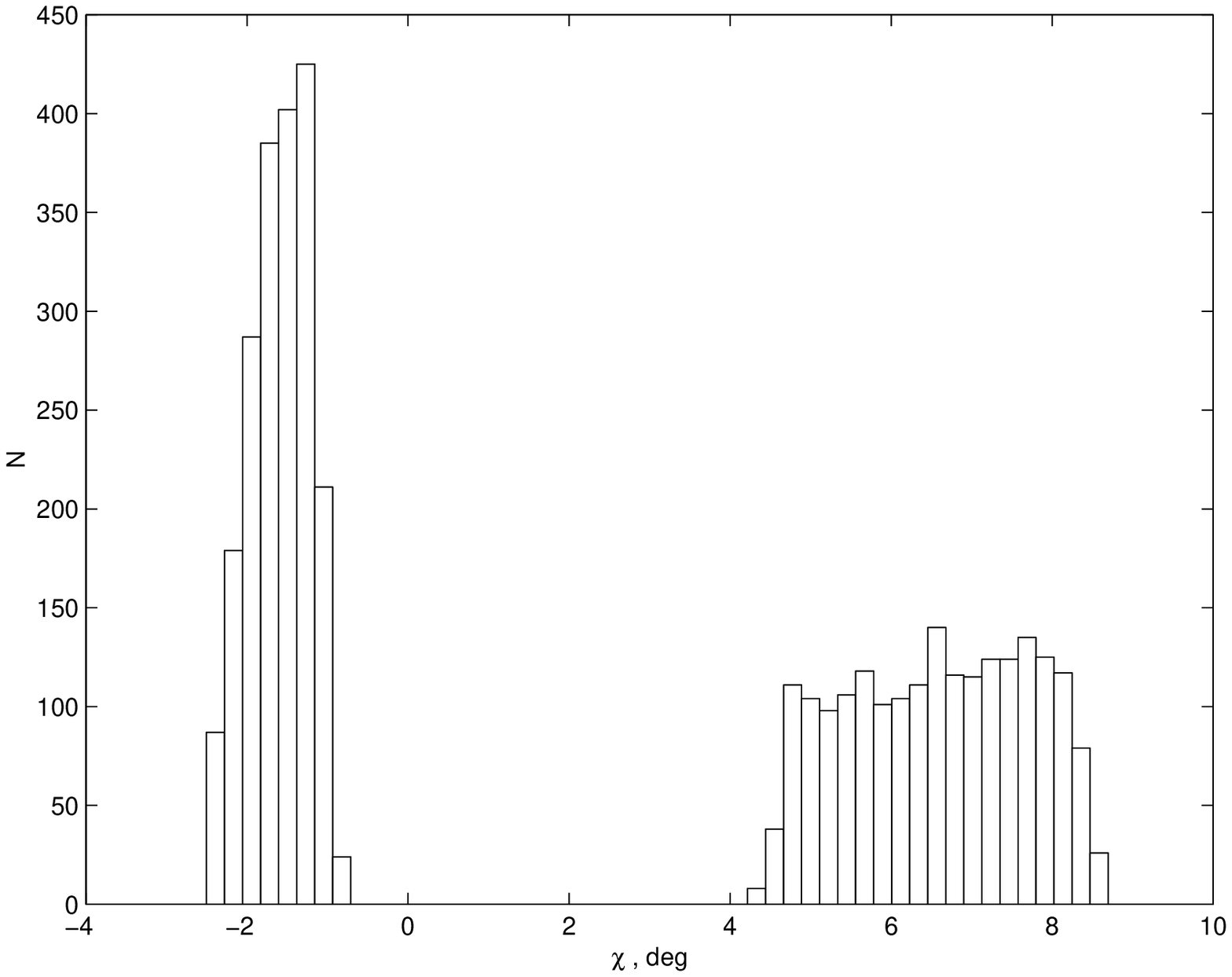}
\includegraphics[width=60mm]{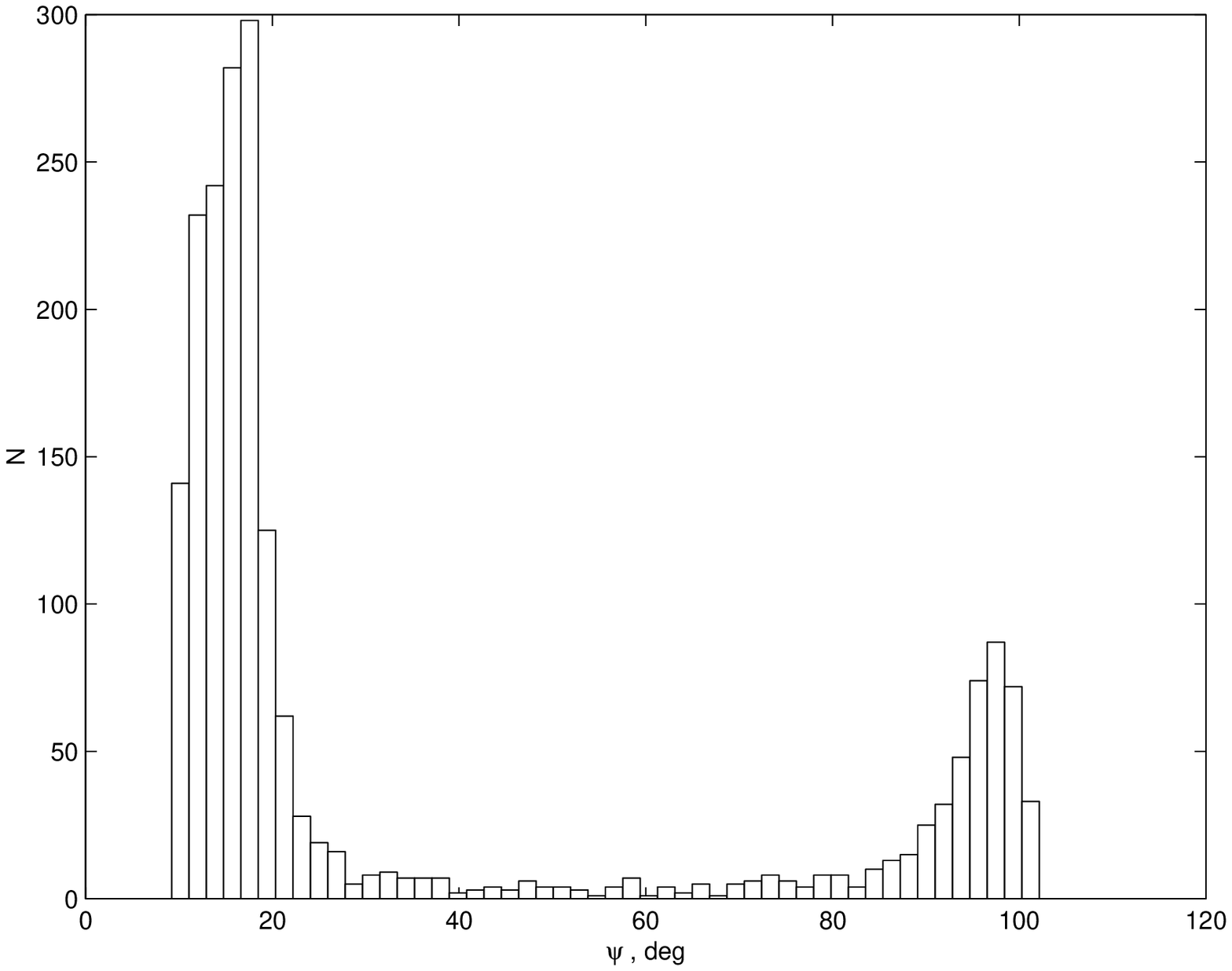}\includegraphics[width=60mm]{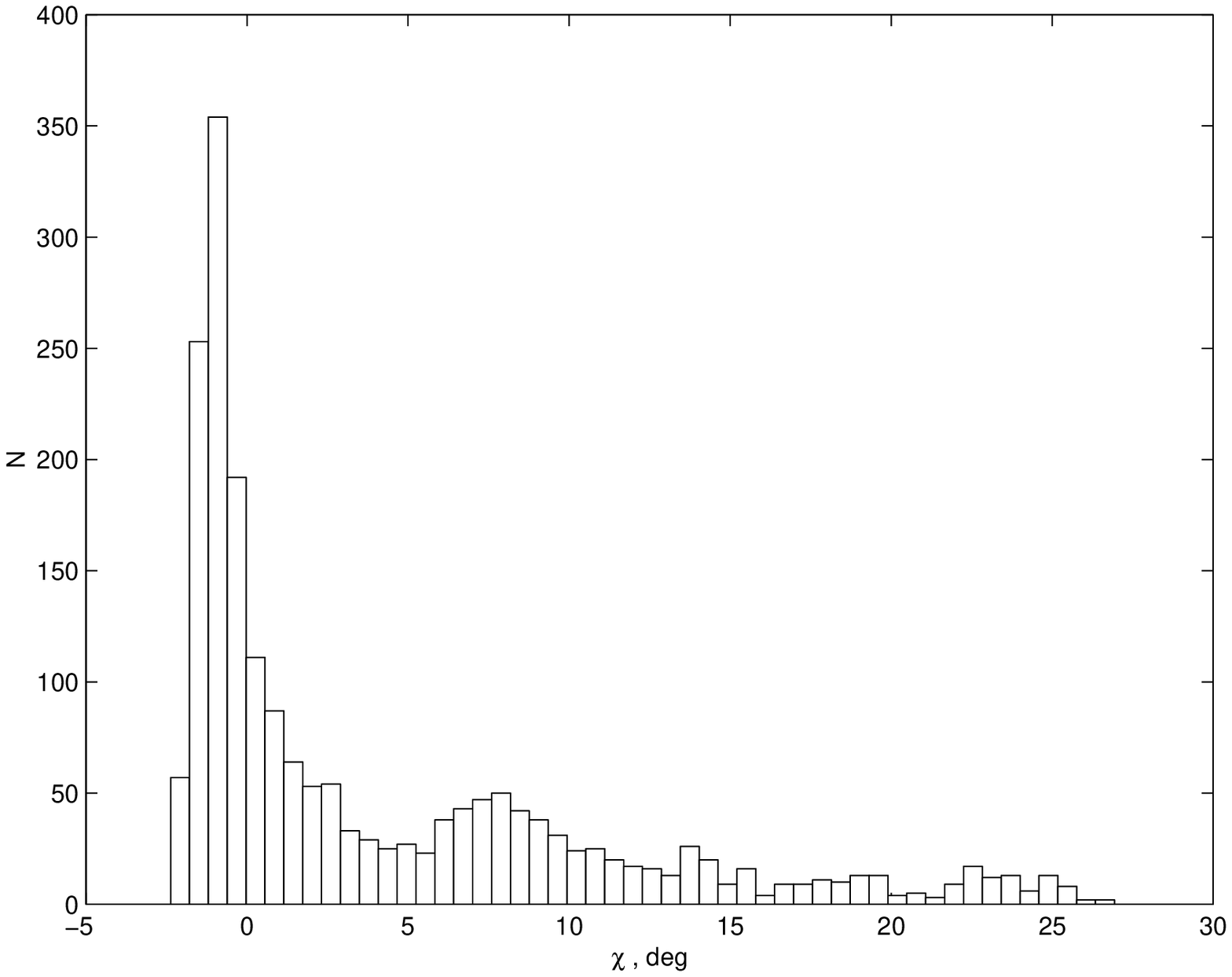}
\includegraphics[width=60mm]{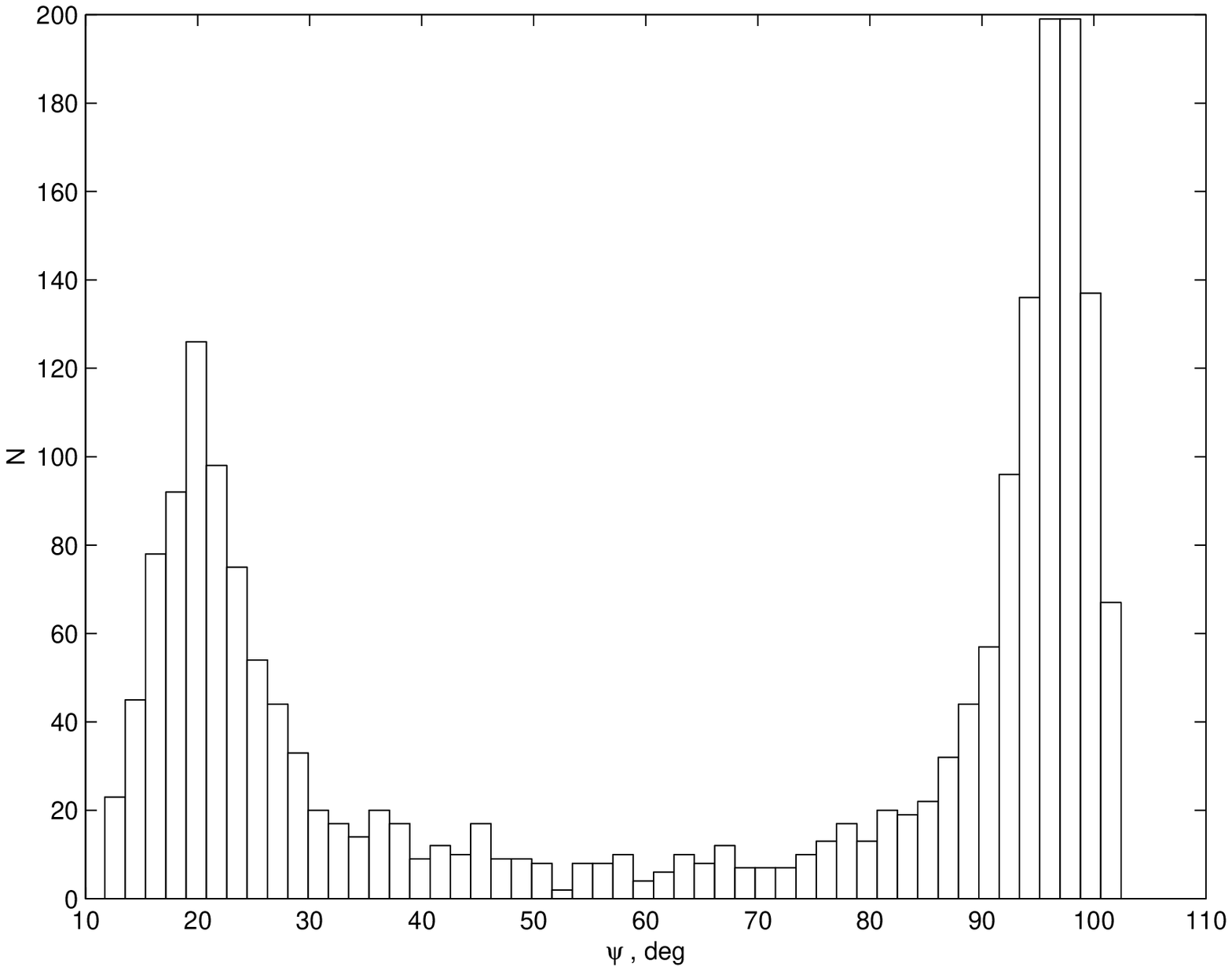}\includegraphics[width=60mm]{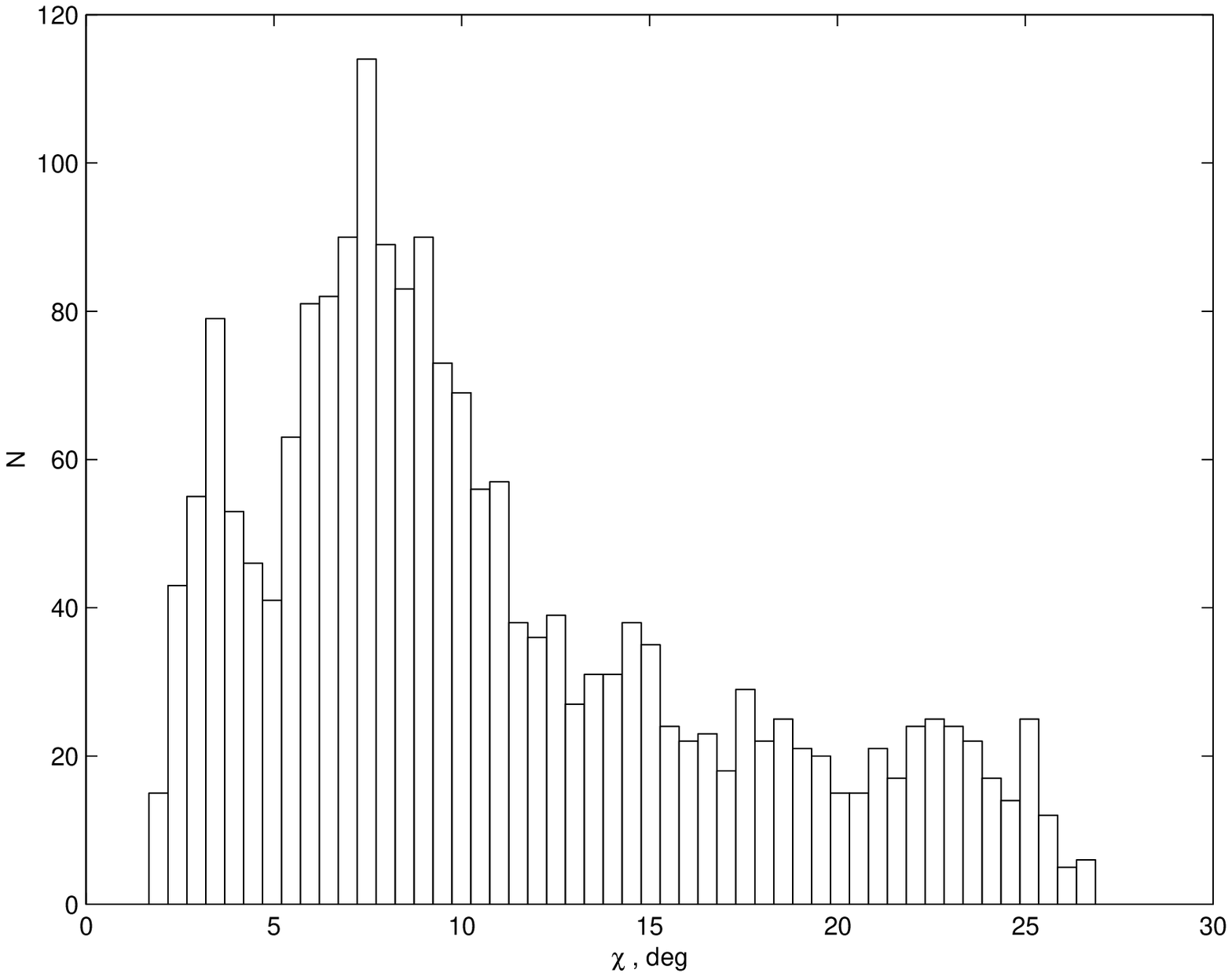}
\caption[]{Numerically simulated histograms of the final PA and
ellipticity after polarization evolution in the fluctuating
plasma. The parameters $\mu$ and $\eta$ are uniformly distributed
over the intervals [0.2,0.4] and [0.9,1.1], respectively. The
upper panel shows the histograms for the original ordinary and
extraordinary modes. The middle and bottom panels correspond to
the sum of modes with random intensity ratio. The coefficient of
conversion is uniformly distributed in the intervals [0,1] and
[0.6,1], respectively.}
\end{figure*}

\begin{figure*}
\includegraphics[width=90mm]{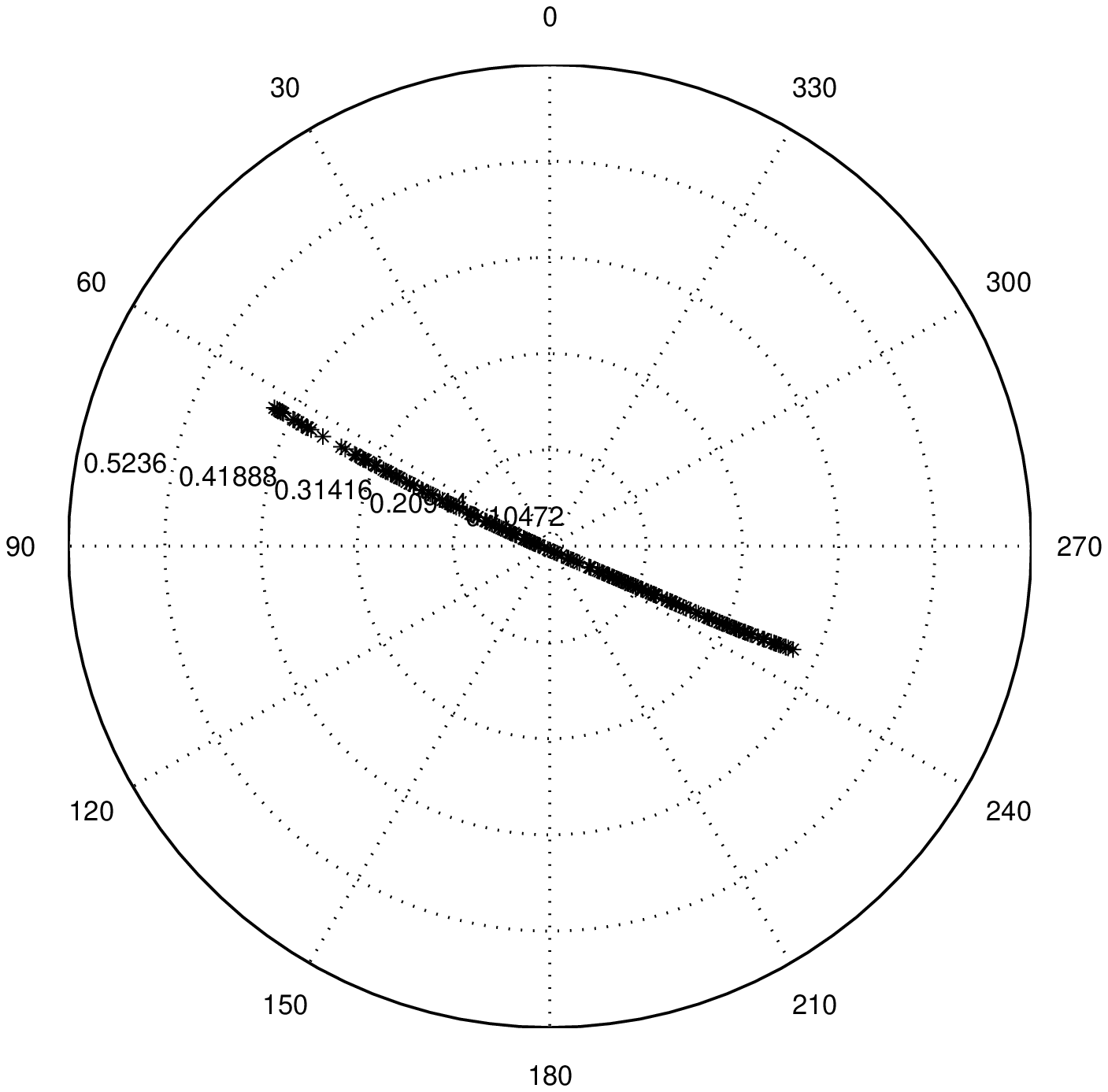}\includegraphics[width=90mm]{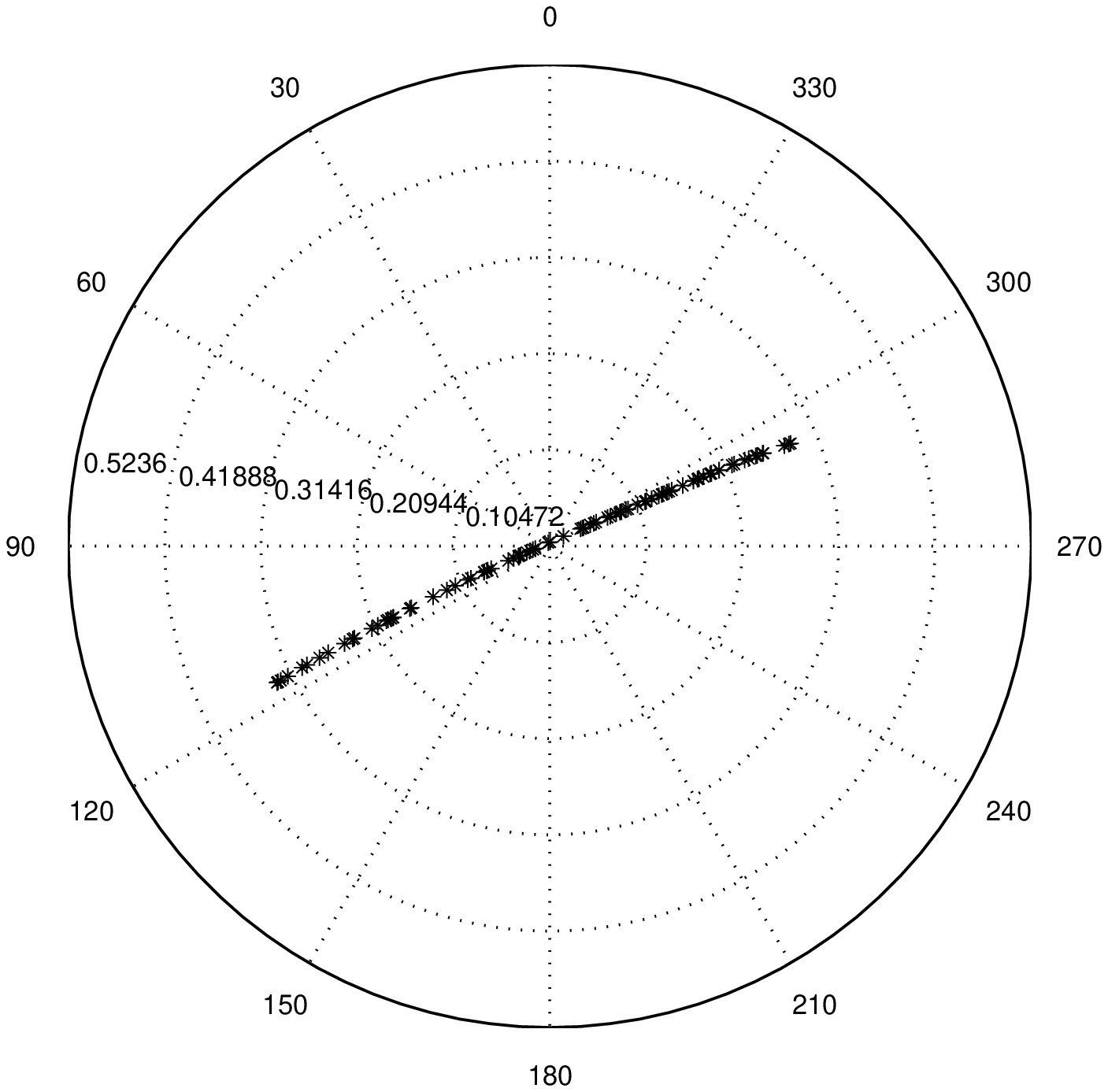}
\caption{Two-dimensional scatter plots of the final Stokes
parameters of the original ordinary (left panel) and extraordinary
(right panel) modes in the Lambert's azimuthal equal area
projection (for more detail see text); $\mu\in [0.1,0.5]$ and
$\eta\in [0.01,0.05]$. The one-to-one correspondence of the
polarization parameters is evident.}
\end{figure*}

\begin{figure*}
\includegraphics[width=90mm]{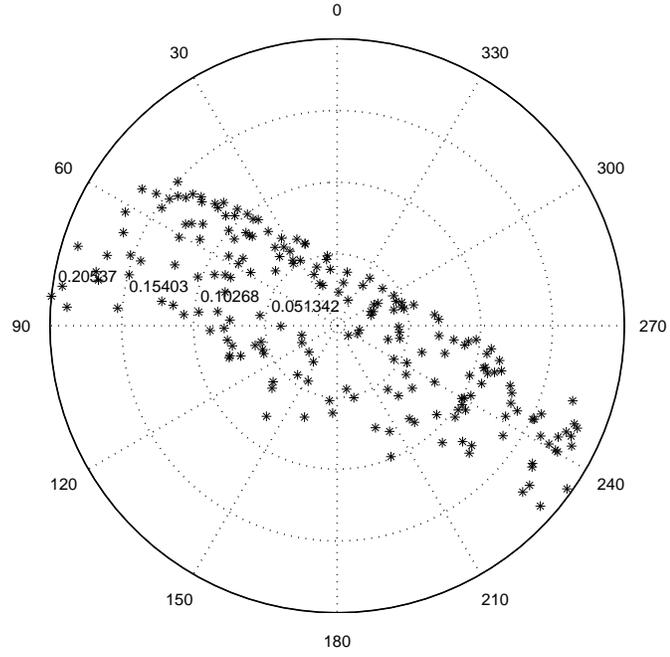}
\caption{The Lambert's azimuthal equal area projection of the
final Stokes parameters of the sum of modes with the dominant
ordinary mode; $\tau\in [0,0.3]$; the role of cyclotron absorption
is not negligible, $\eta\in [0.1,0.4]$; $\mu\in [0.1,0.5]$.}
\end{figure*}

\begin{figure*}
\includegraphics[width=90mm]{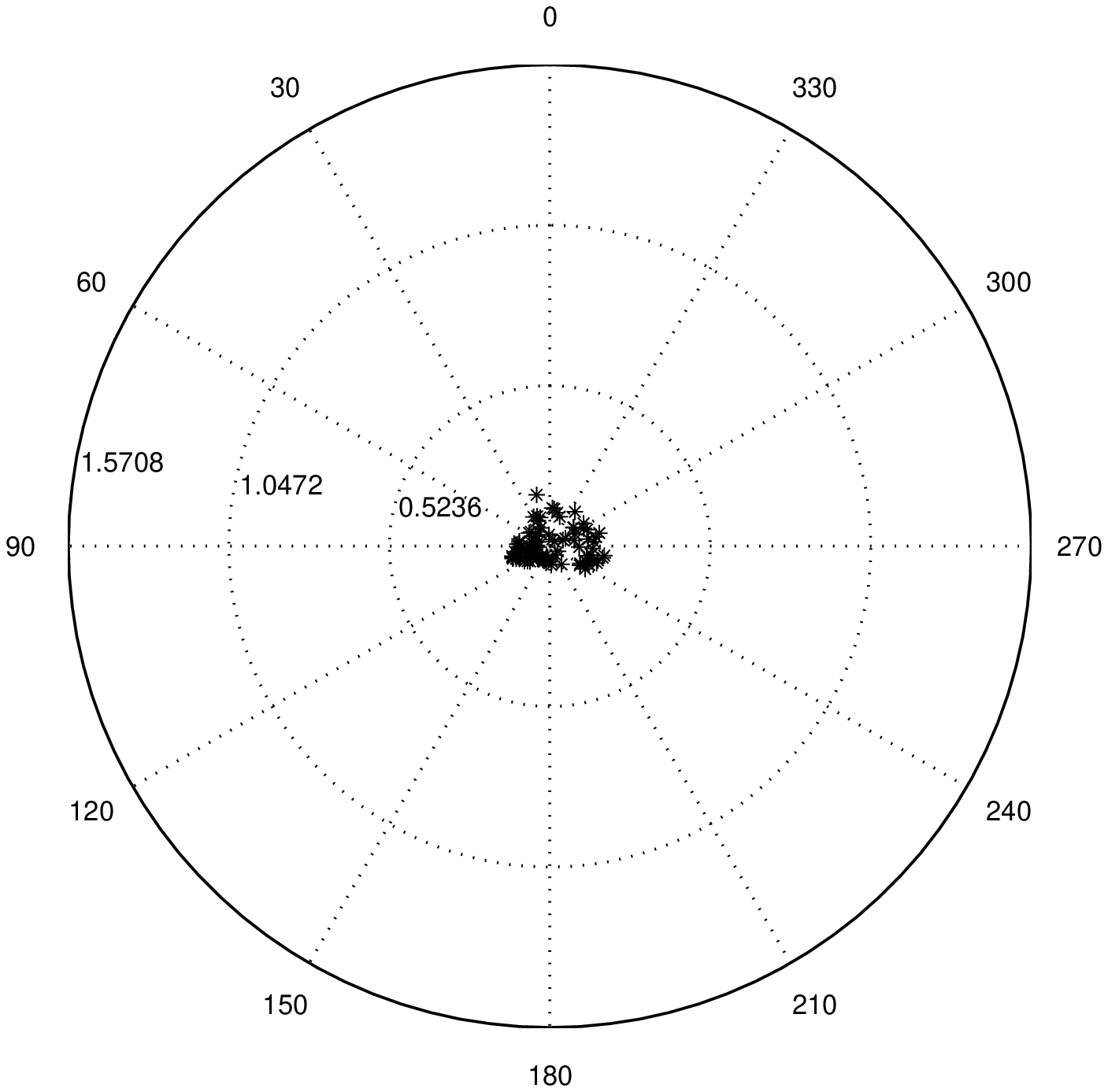}\includegraphics[width=90mm]{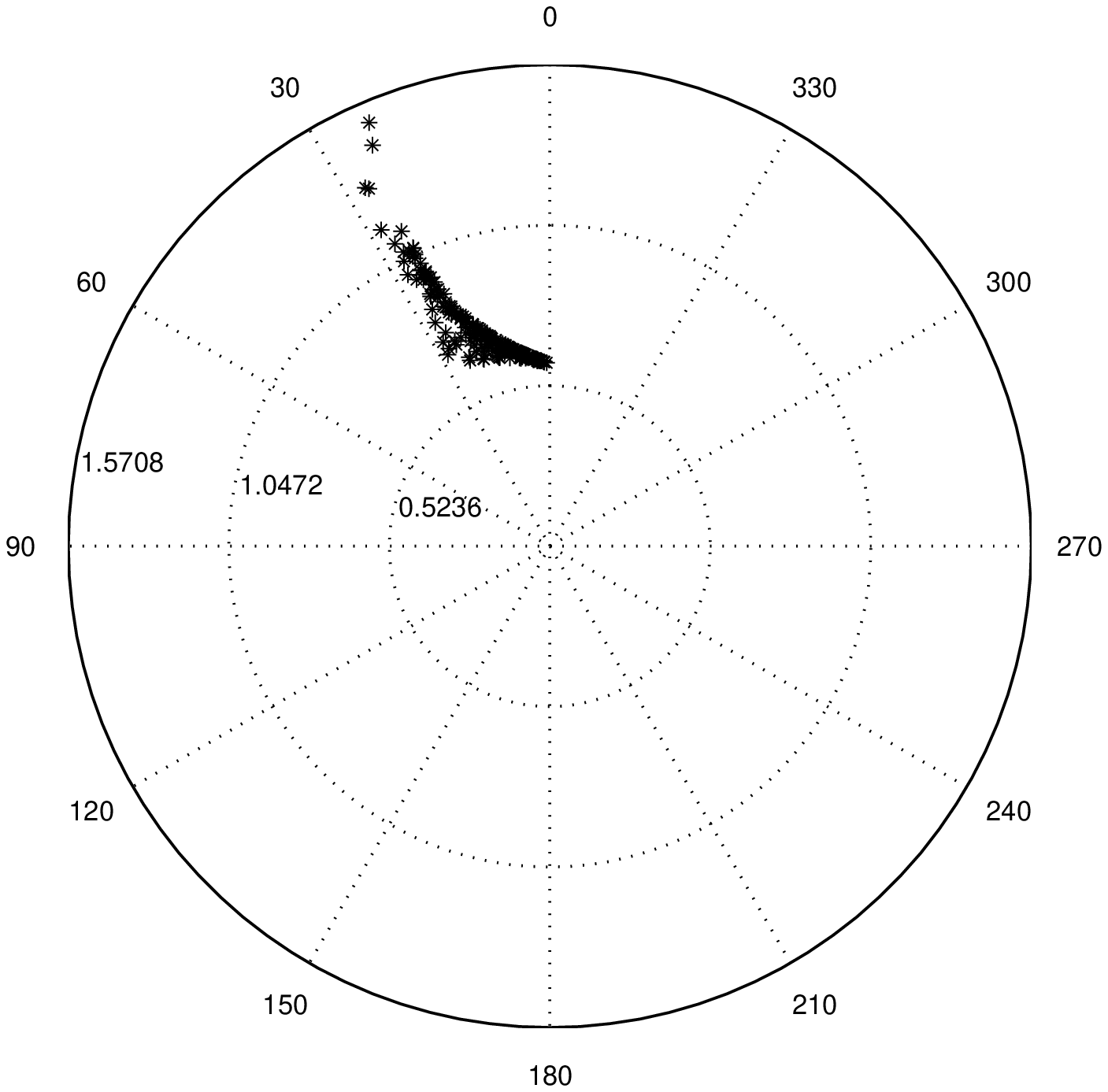}
\includegraphics[width=90mm]{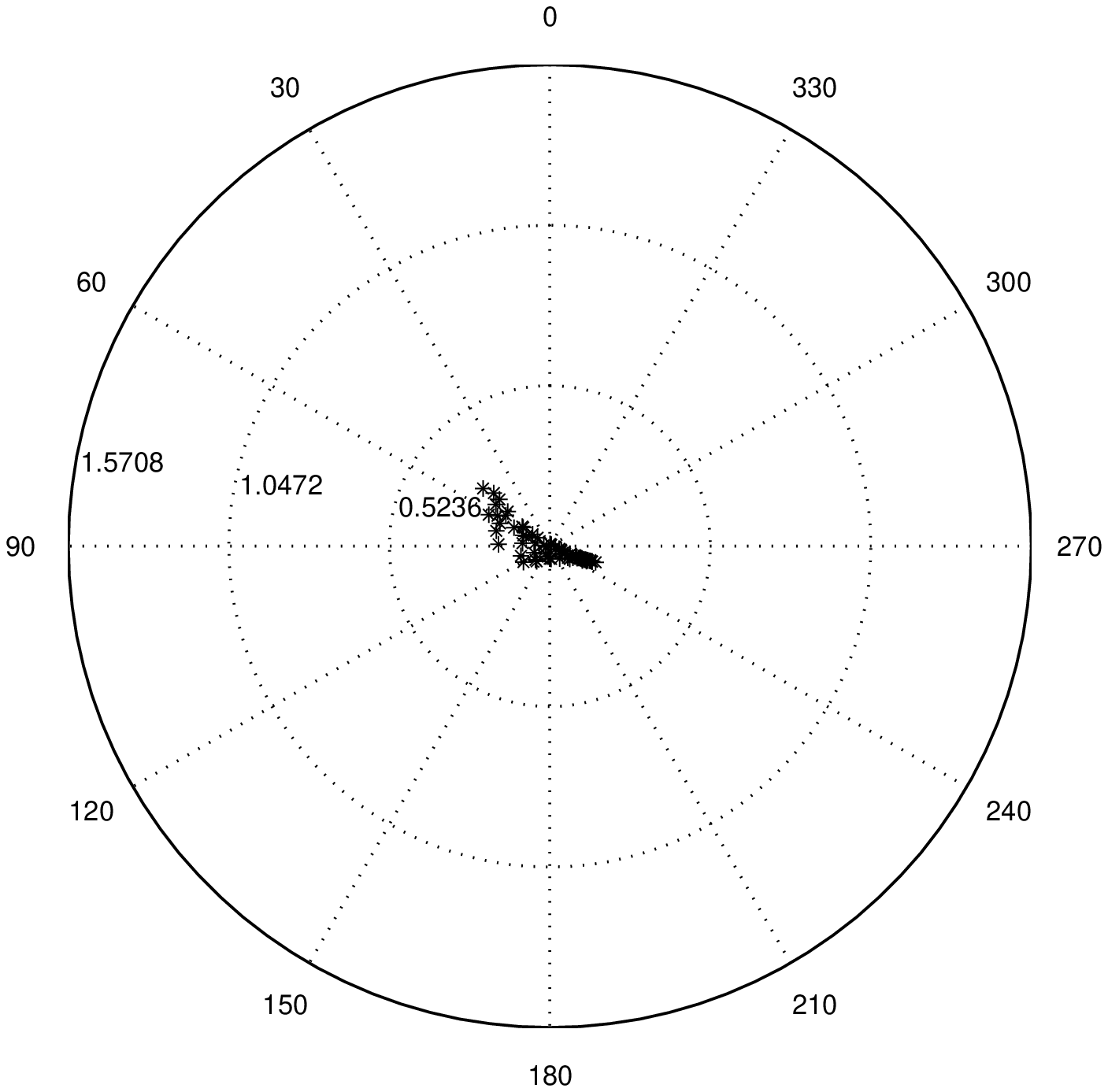}\includegraphics[width=90mm]{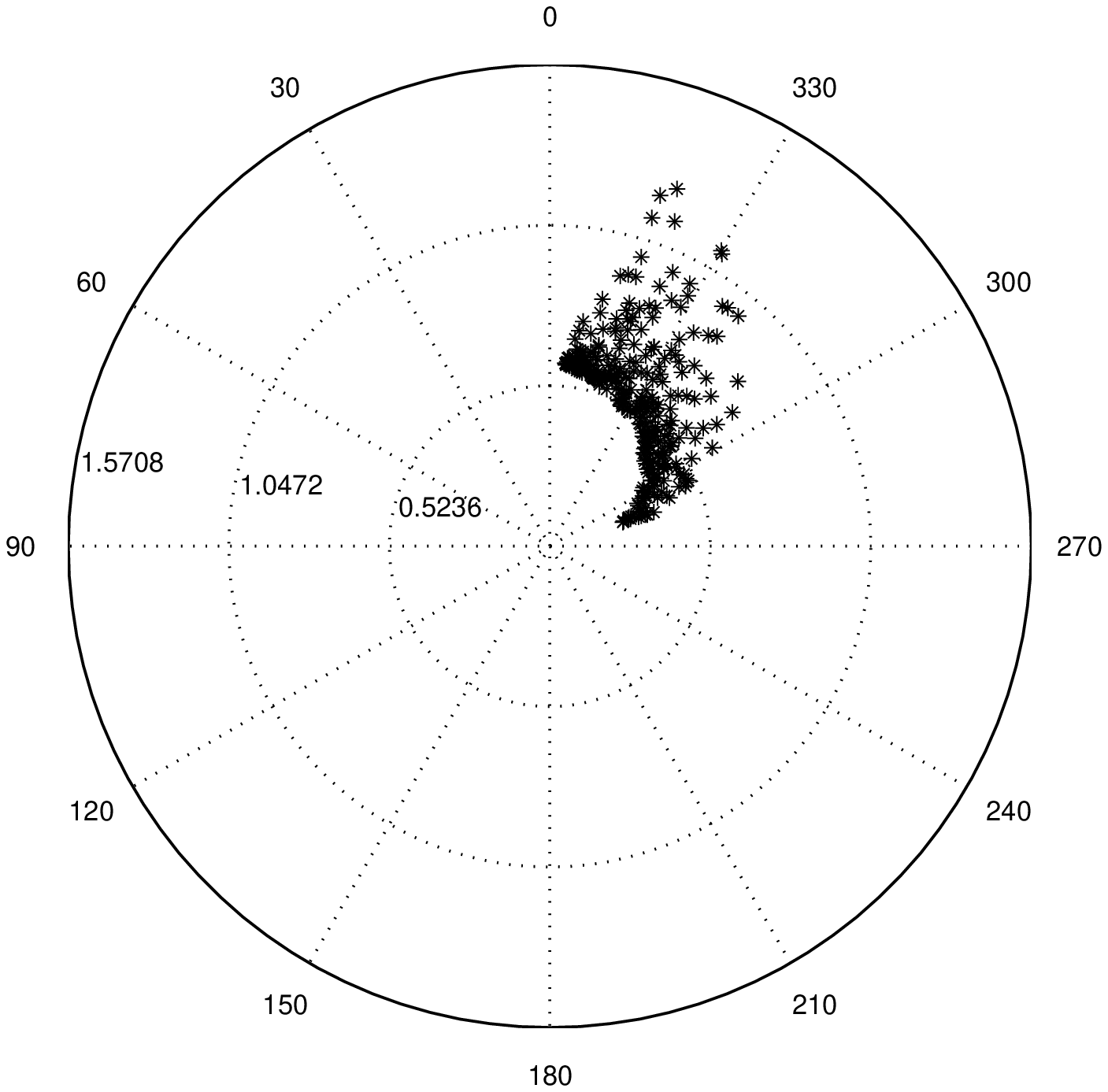}
\caption{Two-dimensional scatter plots of the final Stokes
parameters in case of strong non-orthogonality of the modes,
$\eta\in [0.5,1.5]$. In the upper and lower panels, $\tau\in
[0.5,1]$ and $\tau\in [0,1]$, respectively, and ${\bmath s}_p$ is
centered on the ordinary- and extraordinary-wave positions,
correspondingly. }
\end{figure*}


\end{document}